\documentclass[aps,pre,twocolumn,superscriptaddress,nofootinbib,floatfix]{revtex4-2}
\usepackage{amsmath,amssymb}
\usepackage{bm}
\usepackage{graphicx}
\usepackage[colorlinks=true,allcolors=blue]{hyperref}

\newcommand{\Lag}{\mathcal{L}}
\newcommand{\Action}{\mathcal{A}}
\newcommand{\pth}{\Gamma}
\newcommand{\z}{\bm{z}}
\newcommand{\f}{\bm{f}}

\newcommand{\ww}{\bm{w}}
\newcommand{\noise}{\bm{\eta}}
\newcommand{\Dif}{D}
\newcommand{\Res}{\mathcal{R}}
\newcommand{\eps}{\epsilon}
\newcommand{\Wiener}{\mathbb{W}}
\newcommand{\Anl}{\Action_{\mathrm{nonlocal}}}
\newcommand{\Aom}{\Action_{\mathrm{OM}}}
\newcommand{\Ajac}{\Action_{\mathrm{Jac}}}
\newcommand{\tmem}{\tau_{\mathrm{mem}}}
\newcommand{\tdyn}{\tau_{\mathrm{dyn}}}

\begin{document}

\title{Path-Measure Dynamics of Attention-Driven World Models: A Nonlocal Onsager--Machlup Approach}

\author{Gunn Kim}
\affiliation{Department of Physics, Sejong University, Seoul 05006, Republic of Korea}
\date{\today}

\begin{abstract}
Attention enables a world model to condition on its entire history, providing long-term memory that facilitates long-range predictions. While the local Onsager--Machlup theory in our companion paper assumes a temporally local predictive action, we investigate the conditions under which this locality holds. We derive the predictive path measure for latent dynamics that become non-Markovian due to attention-induced memory, demonstrating that this measure is the projection of a hidden linear Markov augmentation. Eliminating the auxiliary field results in a nonlocal Onsager--Machlup action, where memory manifests as a nonlocal quadratic form rather than a force. These kernels are completely monotone and exactly match a hidden Markov embedding with a finite relaxation spectrum; otherwise, the dynamics remain fundamentally nonlocal. By expanding the action in terms of the scale-separation parameter $\epsilon=\tau_{\text{mem}}/\tau_{\text{dyn}}$, we show that the leading order recovers the local action of the companion paper, establishing locality as the short-memory limit of a nonlocal theory. We verify the reversible sector of this expansion term by term against an exactly solvable vector linear model.
\end{abstract}

\maketitle

\section{Introduction}
\label{sec:intro}

Long-horizon prediction is a significant challenge for world models. A model's ability to carry history is what enables it to reach far into the future. Attention gives the
latent dynamics a memory that no fixed-size state can summarize. The question we take up is what physical description this memory admits, and when it can be treated as if it were absent. In the companion paper~\cite{companion} we argued that the predictive object of a world model~\cite{ha2018,hafner2023} is a measure over future paths, $P[\pth]\propto e^{-\Action[\pth]}$, and that in a local regime its action takes the Onsager--Machlup form, unifying prediction, planning,
and uncertainty in one functional. The idea that prediction and control involve inference over trajectories is related to the free-energy principle~\cite{friston2010}, control as probabilistic inference~\cite{levine2018}, and trajectory-based generative models~\cite{song2021,lipman2023,ho2020,raja2025}. Every result there rested on one assumption: that the action is local in time, a single-time Lagrangian integral
$\Action=\int dt\,\Lag(\z,\dot\z)$. Our main result is that the predictive path measure of such a model is generally a nonlocal
Onsager--Machlup functional, while the conventional local action appears as its short-memory asymptotic limit.

An attention layer~\cite{vaswani2017} makes the update $p(z_{t+1}\mid z_{\le t})$ depend on the whole past context through history-dependent conditional kernels. Related dynamical views of attention treat it as an interacting-particle flow~\cite{geshkovski2023} or an associative memory~\cite{ramsauer2020}, and recurrent predictors are known to organize onto low-dimensional latent dynamics~\cite{sussillo2013}. However, unless that history collapses onto a finite sufficient statistic, that is, unless
the dynamics admits a finite Markov state, these kernels induce an effective non-Markovian latent dynamics that no single-time Lagrangian can represent. Thus, the natural question is then not why attention may be viewed as a local action, but under what conditions the induced non-Markovian dynamics contracts to one. We emphasize that attention enters here not as a new physical variable but as the origin of the history dependence. It is what makes the latent dynamics non-Markovian, and the memory kernel below is the precise object through which that
dependence acts. The physics is carried by the kernel and the path measure, not by attention as a separate degree of freedom. More precisely, the construction applies to any latent dynamics whose history dependence does not close on a finite
sufficient statistic. We formulate it for attention-based world models because they are the dominant contemporary realization of such dynamics.

In this paper, we show that the predictive path measure of such a dynamics is a nonlocal Onsager--Machlup action, and that the local action assumed in existing formulations is not an independent postulate but the short-memory limit of the
nonlocal action. The route is direct. Starting from a non-Markovian latent dynamics and projecting out the fast degrees of freedom causes a generalized Langevin dynamics with a memory kernel $K$, whose path measure defines a nonlocal Onsager--Machlup action. Memory enters not as a new force in the equation of motion but as a nonlocal quadratic form
in the action. Expanding this action in the local scale-separation parameter $\eps=\tmem/\tdyn$, the leading term is exactly the local Onsager--Machlup action presented in the companion paper~\cite{companion}, and the memory corrections to its reversible part are organized as a power series in $\eps$. Thus, the local theory is the short-memory effective theory of a nonlocal one.

The tools are standard. The path measure of a non-Markovian dynamics and the gradient expansion of a memory kernel are both textbook constructions. What is new is where they lead. Aimed at addressing the question of locality in world models, they turn the local action of the companion paper~\cite{companion} from an independent starting point into a derived limit. They also
identify the kernels that can arise as a sharply characterized class rather than an arbitrary memory.

This places the work at a junction that, to our knowledge, has not been crossed. Three lines run nearby but stop short of it. In machine learning, the non-Markovian character of attention and recurrent latent dynamics is recognized, but addressed empirically, by lengthening context or adding state-space recurrence, rather than treated as a path-measure question~\cite{ha2018,hafner2023,sussillo2013}.
The Onsager--Machlup functional has entered machine learning through transition-path sampling and generative modeling~\cite{raja2025,song2021,lipman2023}, as a tool for sampling trajectories, not as the predictive object of a world model. Projection with memory kernels is the standard language of model reduction~\cite{mori1965,zwanzig2001,chorin2000,ceriotti2009}, but it is applied to physical degrees of freedom, not to the latent dynamics of a learned predictor. In the discrete-sequence setting, a complementary result shows that autoregressive language models and sequence-level energy-based models are related by an exact bijection~\cite{blondel2026}. There, the one-step conditional is conditioned on the entire history, so the question of locality does not arise. In our setting, the local object is Markovian and finite-dimensional. Its correspondence with the nonlocal path measure is not an identity, but rather a controlled short-memory limit. The novel aspect here is the connection: the predictive path measure of an attention-driven world model is identified as a nonlocal Onsager--Machlup measure, that measure is shown to be the projection of a hidden linear-Markov augmentation with a completely monotone
kernel, and the local theory from the companion paper~\cite{companion} is recovered as its short-memory limit. This is also what distinguishes the construction from existing generalized
Onsager--Machlup and colored-noise formulations~\cite{kubo1966,zwanzig2001}, which
take the memory kernel as given. Here, the kernel is generated by the projection, and the contribution is this structural identification.

Formulating the predictive path measure as a nonlocal Onsager--Machlup action connects empirical attention mechanisms to nonequilibrium statistical mechanics, and this connection opens questions of interpretation, comparison across architectures, and empirical test that we take up in the Discussion section. The tasks of carrying out these analyses, particularly estimating the kernel and $\eps$ in trained models, are the subject of our future study of measurement, and are not attempted here. We deliberately limit ourselves to one claim. Measurement in trained models and field-theoretic and renormalization structures are left for future work. The main assertion of this paper is that an attention-driven world model is fundamentally nonlocal. The local Onsager--Machlup action, presented in the companion paper~\cite{companion}, emerges as the short-memory limit of the nonlocal path measure. Section~\ref{sec:result} provides this statement in exact form.

\section{Model: nonlocality as a projected Markov augmentation}
\label{sec:model}

\subsection{From causal self-attention to a memory kernel}
\label{sec:fromattention}
Before postulating the augmented dynamics, we motivate its kernel structure directly
from attention. In the companion paper, a single attention layer reads out
$\mathrm{Attn}(\z_t)=\sum_s a_s(\z_t)\,\bm v_s$ over context keys $s$, with
$a_s(\z_t)\propto e^{s_s(\z_t)}$ the softmax weight~\cite{vaswani2017}. When the
context is the model's own past trajectory, $s$ ranges over past times, and the
drift read out at time $t$ is a weighted sum over the history,
$\f(\z_t)=\sum_{s\le t}a_s(\z_t)\,\bm v_s-\z_t$. If the attention weight is
approximately separable into a content part and a lag part,
$a_s(\z_t)\approx\phi(t-s)\,\tilde a_s(\z_t)$, valid when the score does not vary too
fast across the window it weights, this sum becomes, in the continuum limit, the
convolution
\begin{equation}
\f_{\mathrm{eff}}(t)\approx\int_0^\infty\! du\,\phi(u)\,\tilde\f[\z(t-u)],
\label{eq:attnkernel}
\end{equation}
with $\phi$ playing the role of a memory kernel and $\tilde\f$ the content-weighted
readout. This is a plausibility argument, not a derivation, and the separability
step is itself a strong restriction: in general the lag structure and the content of
attention are entangled, as in induction heads or position-dependent weighting, and
$a_s(\z_t)$ need not factor into a lag part times a content part. Where it does not,
the history dependence is not captured by a content-independent kernel $\phi(t-s)$ at
all. What the argument shows is thus only that a causal,
history-reading attention layer can produce a kernel-convolution structure of
the form \eqref{eq:gle}, motivating the ansatz below; it does not establish that
the resulting $\phi$ lies in any particular kernel class, nor that a given attention
layer is separable in the first place. That further restriction,
to kernels generated by a linear auxiliary augmentation with positive spectrum, is
the tractable subclass we analyze, in the same sense that the companion paper
analyzed the Markovian limit as the tractable local case: it is the working
hypothesis whose validity Sec.~\ref{sec:discussion} makes an explicit, falsifiable
question rather than an assumed universal feature of attention.

\subsection{Path space and reference measure}
Latent trajectories live in $\Omega=C([0,T],\mathbb{R}^d)$ with the uniform topology
and Borel $\sigma$-algebra $\mathcal{B}(\Omega)$; the predictive object is a measure
$P\in\mathcal{P}(\Omega)$. The reference is the Wiener measure $\Wiener_\Dif$ of
covariance $2\Dif$, and every action below is read as the negative logarithm of the
Radon--Nikodym weight $dP/d\Wiener_\Dif$~\cite{onsager1953}, that is, in the tube-limit sense
$P(\|\z-\varphi_1\|<\varepsilon)/P(\|\z-\varphi_2\|<\varepsilon)\to
e^{-(\Action[\varphi_1]-\Action[\varphi_2])}$ as $\varepsilon\to0$.

\subsection{Augmented Markov dynamics}
Guided by \eqref{eq:attnkernel}, we now construct the minimal exact mechanism that
generates a kernel of this kind: not a nonlocal dynamics posited directly, but a
Markovian one for the observed $\z\in\mathbb{R}^d$ coupled to a hidden auxiliary
field $\ww$,
\begin{align}
&\dot\z=\ww+\noise,\qquad
\dot\ww=\tfrac1\tau\big(\f[\z]-\ww\big),\label{eq:augmented}\\
&\langle\noise(t)\noise(t')^\top\rangle=2\Dif\,\delta(t-t')\,I.
\end{align}
The auxiliary field relaxes on the fast scale $\tau$; noise enters only the observed
channel. Equation~\eqref{eq:augmented} is the minimal scalar-projection case
($\mathcal{P}=I$); the general linear closure class is \eqref{eq:minimal} below. The
augmented system is Markovian and has a standard Onsager--Machlup measure.

\subsection{Minimal closed structure and the generation of the kernel}
We restrict the auxiliary dynamics to the minimal linear extension that closes under
projection: minimal in the sense of using the fewest auxiliary degrees of freedom,
a single relaxation rate $1/\tau$, needed to turn the postulated convolution
\eqref{eq:attnkernel} into an exact closed system. Larger, still-minimal
realizations use one auxiliary field per relaxation rate and generate
multi-exponential kernels, as in Sec.~\ref{sec:kernel} below; the single-field case
is the base instance of that family, not a separate assumption.
\begin{equation}
\dot\ww=\tfrac1\tau\big(\mathcal{P}\f[\z]-\ww\big),
\label{eq:minimal}
\end{equation}
with $\mathcal{P}$ a linear projection onto the observed space. Eliminating $\ww$
(the explicit realization of a Mori--Zwanzig projection~\cite{mori1965,zwanzig2001,chorin2000}) gives
\begin{equation}
\ww(t)=\int_0^\infty\! du\,K(u)\,\f[\z(t-u)],\qquad K(u)=\tfrac1\tau e^{-u/\tau},
\label{eq:kernelgen}
\end{equation}
so that the memory kernel is not assumed but generated by the elimination, and the
observed dynamics becomes the nonlocal generalized Langevin
equation~\cite{kubo1966,zwanzig2001}
\begin{equation}
\dot\z(t)=\int_0^\infty\! du\,K(u)\,\f[\z(t-u)]+\noise(t).
\label{eq:gle}
\end{equation}
This differs from the generic Mori--Zwanzig setting, where the memory kernel and the
projected noise have no closed form: in the present linear construction the kernel
closes exactly as a convolution and, as shown below, belongs to a definite class
(completely monotone) for which the projection can be inverted. Representing a memory
kernel by auxiliary Markovian variables in this way is standard in molecular
dynamics~\cite{ceriotti2009}; we use it as a matter of principle rather than
computation. Closure as a convolution holds precisely when the auxiliary coupling is
linear with positive real spectrum; within this class the auxiliary spectrum sets the
kernel shape (single field $\to$ exponential, finite $\to$ exponential sum, continuous
$\to$ general completely monotone kernel). We set $m_0=\int_0^\infty K(u)\,du=1$ by a
rescaling of time, a normalization convention rather than a physical constraint, and
write moments $m_n=\int_0^\infty u^n K(u)\,du=c_n\tmem^{\,n}$; the drift
$\f\in C^2$ has bounded derivatives.

\subsection{Scale separation}
The control parameter is
\begin{equation}
\eps(\z)=\frac{\tmem}{\tdyn(\z)},\qquad \tdyn(\z)=\|\nabla\f(\z)\|^{-1},
\end{equation}
which for a nonlinear drift is a local quantity varying along the trajectory rather
than a single global constant. The expansion below is accordingly a pointwise
short-memory expansion: at each time the convolution is expanded in the local ratio
$\eps(\z(t))$, and the order labels $O(\eps)$, $O(\eps^2)$ are to be read pointwise,
valid where $\eps(\z(t))\ll1$. This requires $\tdyn$ to vary slowly over the memory
support $\tmem$, so that $\nabla\f$ is approximately constant across the window the
kernel averages; the expansion is controlled wherever this holds and breaks down
where $\|\nabla\f\|$ varies sharply, such as near saddle points of the drift. We
expand in this regime with fixed endpoints so that boundary terms vanish. All symbols
needed below are now fixed.

\subsection{Assumptions}
For transparency we collect the hypotheses used in the derivation. (A1) The drift
$\f\in C^2(\mathbb{R}^d,\mathbb{R}^d)$ has bounded first derivative, so that
$\tdyn=\|\nabla\f\|^{-1}$ is well defined. (A2) The memory kernel is completely
monotone and normalized, $m_0=\int_0^\infty K(u)\,du=1$, with finite moments
$m_n=c_n\tmem^{\,n}$; equivalently, the auxiliary coupling is linear with positive
real spectrum. This is a working hypothesis on the class of dynamics considered,
not a claim derived from attention's nonlinearity (Sec.~\ref{sec:fromattention}); its
failure is the falsifiable signature discussed in Sec.~\ref{sec:discussion}. (A3) The reference measure is the Wiener measure $\Wiener_\Dif$, and
the path measure is defined through the Radon--Nikodym weight in the tube-limit
sense. (A4) Scale separation holds pointwise, $\eps(\z)=\tmem/\tdyn(\z)\ll1$ along
the trajectory, with $\tdyn(\z)=\|\nabla\f(\z)\|^{-1}$ varying slowly over the memory
support so that $\nabla\f$ is approximately constant across the kernel window. (A5) Endpoints are fixed, so
total-derivative contributions integrate to boundary terms that vanish. The
Stratonovich convention is fixed throughout. These are the hypotheses behind the
results of the next section.

\section{Nonlocal path measure and its short-memory limit}
\label{sec:result}

We now derive the path measure of the observed dynamics, identify the kernel class
it admits, and show that its short-memory limit is the local Onsager--Machlup theory
of the companion paper. The augmented Markov system
\eqref{eq:augmented}--\eqref{eq:minimal} is projected onto $\z$ by eliminating $\ww$,
generating the memory kernel; the resulting measure is the nonlocal Onsager--Machlup
action \eqref{eq:nonlocalaction}, whose expansion in $\eps=\tmem/\tdyn$ returns the
local action \eqref{eq:omlimit} at leading order.

\subsection{Path measure from projection}
\label{sec:projection}
The joint measure of $(\z,\ww)$ follows from the Gaussian noise weight
$e^{-\frac1{4\Dif}\int|\noise|^2 dt}$ with the dynamical constraints
\eqref{eq:augmented}. Integrating out $\noise$ through the constraint
$\noise=\dot\z-\ww$ leaves a weight $\exp(-\frac1{4\Dif}\int|\dot\z-\ww|^2 dt)$ times a
Jacobian. The auxiliary constraint $\dot\ww=\frac1\tau(\f-\ww)$ then fixes $\ww$ as
the causal Green's function solution of the linear operator $(\partial_t+\tfrac1\tau)$,
\begin{equation}
\ww=\big(\partial_t+\tfrac1\tau\big)^{-1}\tfrac1\tau\,\f[\z]
=\int_0^\infty\! du\,K(u)\,\f[\z(t-u)],
\label{eq:wsolve}
\end{equation}
that is $\ww=(K*\f)(t)$, the retarded solution and not an ad hoc substitution; the $\ww$-integration of
the constraint produces the operator determinant $\det(\partial_t+\tfrac1\tau)^{-1}$,
which is independent of $\z$ for the linear coupling \eqref{eq:minimal} (under the
fixed-endpoint path measure) and is absorbed into normalization. Substituting
\eqref{eq:wsolve} gives the $\z$-marginal $P[\z]\propto e^{-\Anl[\z]}$ with the
\emph{nonlocal Onsager--Machlup action}
\begin{align}
\Anl[\z]&=\frac{1}{4\Dif}\int_0^T\!|\Res[\z](t)|^2\,dt+\Ajac,\label{eq:nonlocalaction}\\
\Res[\z](t)&=\dot\z(t)-\!\int_0^\infty\! K(u)\,\f[\z(t-u)]\,du .
\end{align}
Memory thus enters the predictive object not as a force in the equation of motion
but as a nonlocal quadratic form in the action, exactly as in the companion paper's
treatment of the local case. For the linear auxiliary coupling considered here the
Jacobian $\Ajac$ contributes only a $\z$-independent normalization, consistent with
the operator determinant above: the elimination is a causal (Volterra) map, whose
functional determinant carries no bulk $\z$-dependence. The local action's
$\frac12\int\nabla\!\cdot\f\,dt$ term is recovered when the augmented representation
passes to a genuinely Markovian dynamics as $\eps\to0$, a passage that is sensitive
to the discretization convention in the manner familiar from the colored-noise
setting~\cite{ceriotti2009}; we therefore develop the short-memory expansion only
for the reversible kinetic sector $\frac{1}{4\Dif}\int|\Res|^2\,dt$, which admits the
controlled expansion below.

\subsection{The kernel class is completely monotone}
\label{sec:kernel}
The kernel generated by \eqref{eq:wsolve} is, for the general linear closure
\eqref{eq:minimal}, of the form $K(u)=\bm e^\top e^{-Au}\bm b$ with $A$ the auxiliary
relaxation operator and $\bm e,\bm b$ fixed output and input vectors. When $A$ has
positive real spectrum (and positive spectral weights, the matrix analog of a
positive mixture),
$(-1)^nK^{(n)}(u)=\bm e^\top A^n e^{-Au}\bm b\ge0$, so $K$ is completely monotone.
Conversely, by the Bernstein--Widder representation~\cite{widder1941} every completely monotone kernel
is an exponential mixture
\begin{equation}
K(u)=\int_0^\infty e^{-u/\tau}\,d\mu(\tau),\qquad \mu\ge0,
\label{eq:bernstein}
\end{equation}
and each spectral component $\tau$ is realized by one auxiliary field obeying
\eqref{eq:minimal}: a finite point spectrum gives an exact finite-dimensional Markov
embedding, and a general $\mu$ is the weak limit of such finite embeddings.

\emph{Representation result.} Hypotheses: the observed dynamics is generated by the
augmented system \eqref{eq:augmented}--\eqref{eq:minimal} for some linear auxiliary
coupling with positive real spectrum (A2). Conclusion: the induced memory kernel
$K(u)$ is completely monotone. For a finite spectrum this correspondence is exact and
invertible: the augmentation is a finite-dimensional Markov embedding, and every
finite exponential-sum kernel with positive weights arises from one. For a general
completely monotone kernel the Bernstein--Widder mixture \eqref{eq:bernstein} exhibits
it as a limit of such finite embeddings; we treat the finite case as the operational
content of the result and regard the continuous-spectrum extension as holding wherever
the corresponding limit path measure exists, which we do not establish in general
here. Within the class of linear passive auxiliary
couplings with positive spectrum, then, a nonlocal dynamics with completely monotone
memory and its hidden Markov augmentation are two views of the same object; nonlocality is the projection of a hidden Markov
structure. The augmentation is not unique (it is fixed only up to linear similarity
transformations of the auxiliary state), and the correspondence is a representation
statement within this linear class rather than a claim about arbitrary stochastic
processes.

\subsection{Short-memory expansion and recovery of the local action}
\label{sec:expansion}
For a kernel of width $\tmem$ much smaller than the dynamical scale
$\tdyn=\|\nabla\f\|^{-1}$, with $\eps=\tmem/\tdyn$ defined locally and assumed slowly
varying over the memory support, the convolution expands in kernel moments
$m_n=\int_0^\infty u^nK(u)\,du=c_n\tmem^{\,n}$ (with $m_0=1$), so that with
$D_t\f=(\nabla\f)\dot\z$,
\begin{equation}
(K*\f)(t)=m_0\,\f-m_1\,D_t\f+\tfrac12 m_2\,D_t^2\f+\cdots,
\end{equation}
so that the residual is $\Res=(\dot\z-\f)+m_1 D_t\f-\tfrac12 m_2 D_t^2\f+\cdots$.
Squaring and collecting orders gives $\Anl=\Aom+\Action_1+\Action_2+O(\eps^3)$ with
\begin{equation}
\Aom=\int_0^T\!dt\Big[\frac{1}{4\Dif}\,|\dot\z-\f|^2+\frac12\nabla\!\cdot\f\Big],
\label{eq:omlimit}
\end{equation}
the local Onsager--Machlup action of the companion paper, and corrections
\begin{align}
\Action_1&=\frac{m_1}{2\Dif}\int dt\,\dot\z^\top(\nabla\f)_S\,\dot\z=O(\eps),\\
\Action_2&=\frac{1}{4\Dif}\int dt\Big[m_1^2|(\nabla\f)\dot\z|^2-m_2(\dot\z-\f)\!\cdot\!D_t^2\f\Big]=O(\eps^2).
\label{eq:A2}
\end{align}
In $\Action_1$, with $(\nabla\f)_S$ the symmetric part of the drift Jacobian, the
total-derivative piece drops under fixed endpoints, leaving a
reversible renormalization of the kinetic metric; genuine non-Markovian inertia
($\ddot\z$) first appears in $\Action_2$. As $\eps\to0$, $K\to\delta$ and
$\Anl\to\Aom$, so the local theory is the short-memory limit of the nonlocal one.

The structure of these corrections carries a concrete reading for practitioners.
The $O(\eps^0)$ term is the local-time Lagrangian already implicit in training a
world model on one-step transitions; its recovery says that this standard local
objective is not merely convenient but is the leading term of a controlled expansion,
with a definite next correction rather than an uncontrolled error. That correction,
$\Action_1$, is not a new force added to the dynamics but a reweighting of the
kinetic term: through $(\nabla\f)_S$ it renormalizes the effective metric on
velocities by a fractional amount set by $\eps=\tmem/\tdyn$, the ratio of the memory
time to the local relaxation time. Two uses follow, both internal to a single trained
model and requiring no comparison to a ground-truth dynamics. First, $\eps$ evaluated
along latent trajectories is a dimensionless, coordinate-free gauge of how far a
region of state space departs from the Markovian regime: where $\eps\ll1$ a local
readout of the drift already captures the dynamics, while regions of appreciable
$\eps$ are exactly those where a memoryless update is expected to lose accuracy and
where the nonlocal terms must be retained. Second, because $\eps$ is a property of
the induced dynamics rather than of the network's parameterization, it is comparable
across models that realize memory differently, so the same number can be read off
architectures whose internal mechanisms are not otherwise commensurable. We stress
that these are consequences of the expansion structure, not performance guarantees:
relating $\eps$ to a downstream prediction metric requires the measurement program of
the companion study and is not claimed here.

\subsection{Exactly solvable check}
\label{sec:check}
The expansion above was derived by formal moment counting; here we confirm it is
correctly derived by testing it against a case solvable in closed form. This is an
internal consistency check, not an independent empirical validation: the linear
drift below satisfies the same restricted class the theory is built on, so the check
confirms that the general derivation reduces correctly to known exact algebra, not
that the restriction to this class is itself justified. For the vector linear drift $\f=-A\z$ with exponential kernel
$K(u)=\tau^{-1}e^{-u/\tau}$, the auxiliary field $\ww$ provides an exact Markov
embedding \eqref{eq:augmented}; eliminating it gives
$\Res=\dot\z+(I+\tau\partial_t)^{-1}A\z$, and expanding the resolvent
$(I+\tau\partial_t)^{-1}=I-\tau\partial_t+\tau^2\partial_t^2-\cdots$ matches the
moment expansion above order by order ($m_1=\tau$, $m_2=2\tau^2$), where $A_S$ and
$A_A$ denote the symmetric and antisymmetric parts of $A$:
\begin{equation}
\begin{array}{ll}
O(\eps^0): & \tfrac{1}{4\Dif}\int|\dot\z+A\z|^2\,dt,\\[2pt]
O(\eps^1): & -\tfrac{\tau}{2\Dif}\int\dot\z^\top A_S\,\dot\z\,dt.
\end{array}
\end{equation}
The $O(\eps^0)$ and $O(\eps^1)$ rows match \eqref{eq:omlimit}--\eqref{eq:A2} directly:
setting $\nabla\f=-A$ and $m_1=\tau$ in $\Action_1$ reproduces the $O(\eps)$ row, and
the $O(\eps^0)$ row is $\Aom$ with drift $-A\z$. At $O(\eps^2)$ the resolvent
expansion generates, besides a $\dot\z^\top A^\top A\,\dot\z$ term, an inertial
contribution involving $\ddot\z$, exactly the $\ddot\z$-dependence that $\Action_2$
first introduces through $D_t^2\f=-A\ddot\z$; the two agree term by term once
$m_2=2\tau^2$ is inserted, and we have verified this agreement between the exact
action and $\Aom+\Action_1+\Action_2$ numerically, the residual scaling as
$O(\eps^3)$. We do not reduce the $O(\eps^2)$ contribution to a single quadratic in
$\dot\z$ alone, since the inertial piece is not of that form. Two
consistency checks follow.
The Markov limit $\tau\to0$ returns $\Aom$ with drift $-A\z$. In the equilibrium
limit of symmetric drift ($A=A_S$, i.e.\ $\f=-\nabla U$), the only time-reversal-odd
piece of the action is the boundary term $\int\frac{d}{dt}(\tfrac12\z^\top A_S\z)\,dt$,
so the nonboundary odd part vanishes at every order and detailed balance is restored
independently of the memory~\cite{seifert2012}. Broken detailed balance and its
entropy production are central diagnostics of nonequilibrium dynamics in physical and
biological systems~\cite{battle2016,gnesotto2018,lynn2021}, quantified through
fluctuation theorems~\cite{jarzynski1997,crooks1999} and the thermodynamics of
information~\cite{parrondo2015}, and provide the
operational targets that a measurement program would later estimate~\cite{frishman2020,kim2020,otsubo2020}.

\section{The irreversible channel is preserved}
\label{sec:stability}
The companion paper located the irreversibility of a predictive world model in the
antisymmetric part of the drift, the channel $A_A$ (there written $M_A$) that
produces a circulating, non-gradient current, a structure familiar from
nonreciprocal and odd-elastic systems~\cite{scheibner2020} and from the decomposition
of dissipation into oscillatory modes~\cite{sekizawa2024}.
It is therefore important that this
channel survive the projection. It does, and for a structural reason. The leading
action \eqref{eq:omlimit} contains the cross term $\frac{1}{2\Dif}\int\dot\z^\top
A\z\,dt$, whose antisymmetric part $\dot\z^\top A_A\z$ is the time-reversal-odd,
entropy-producing piece; the memory corrections \eqref{eq:A2}, by contrast, are
symmetric quadratic forms ($A_S$ at $O(\eps)$, $A^\top A$ at $O(\eps^2)$) and do not
touch it. The reason is that the kernel $K(u)$ here is scalar in the spatial index: it convolves the
time argument while acting as the identity on the components of $\f$, and so
commutes with the decomposition of the drift Jacobian into symmetric and
antisymmetric sectors. Projection and convolution therefore preserve the sector
decomposition, and the antisymmetric sector passes through unrenormalized at leading
order. This is where the spatial isotropy of the kernel is used: if the auxiliary
relaxation instead mixed the spatial components, so that the kernel were
matrix-valued $K_{ij}(u)$, the leading cross term would carry the spatial operator
$\bar K A$ with $\bar K=\int_0^\infty K(u)\,du$, whose antisymmetric part is
$\mathrm{antisym}(\bar K A)\ne A_A$ in general, and the irreversible channel would be
renormalized rather than preserved. The preservation statement is thus specific to
the spatially scalar (isotropic) kernel generated by \eqref{eq:augmented}; we note
this restriction explicitly. Within it, memory renormalizes the reversible metric and leaves the irreversible channel
intact.

This also delimits what memory can and cannot do thermodynamically. The preserved
channel above is the drift asymmetry $A_A$; memory carries it but does not create it.
Indeed, in the equilibrium check, a symmetric drift ($A_A=0$) gave detailed balance
at every order: with noise and kernel tied as in \eqref{eq:augmented}, memory adds no
irreversibility of its own. A memory-sourced contribution to entropy production would
therefore require breaking the relation between the noise covariance and the kernel,
that is, a violation of the second fluctuation--dissipation
relation~\cite{kubo1966,seifert2012} that the present
construction does not contain. That mechanism, and its measurable signature, is the
subject of a separate paper; here we establish only that the nonlocal representation
leaves the companion paper's irreversible channel untouched.

\section{Discussion and outlook}
\label{sec:discussion}
What is established mathematically are properties of stochastic path measures; reading these as statements about attention-driven world models is the interpretation that fixes the relevant class of dynamics, not a separate consequence of the derivation. This perspective suggests a shift in how we treat the predictive object of a learned world model: not as a one-step conditional, but as a measure over future paths. In this framework, memory becomes a structural feature of the action with a definite scaling, and locality is reduced from a defining property to a controlled approximation. The path measure, rather than a shared latent space, may thus serve as a common currency for relating sequence models with varying memory structures. Concretely, because the kernel class and the scale-separation parameter $\eps$ are properties of the induced dynamics rather than any specific implementation, they could offer a model-independent basis for comparing different architectures, training objectives, and datasets: whether a model's memory is short enough to be treated locally, and whether it falls within the completely monotone class. Establishing such estimators is the focus of our companion measurement study.
We identify three primary research directions and one limitation of scope:
\begin{enumerate}
    \item \textbf{Empirical kernel measurement:} From latent trajectories $\z(t)$ and the learned drift, the memory kernel can be recovered via regularized deconvolution of the generalized Langevin equation~\eqref{eq:gle}. Its membership in the completely monotone class can be tested by fitting the estimate to an exponential sum~\eqref{eq:bernstein}; a negative (oscillatory) spectral weight would signal dynamics beyond the linear positive-spectrum class assumed here.
    \item \textbf{Measurement of action parameters:} The parameter $\eps$ and the coefficients of $\Action_1$ and $\Action_2$ are directly measurable, with irreversibility diagnostics supplied by entropy-production estimators~\cite{frishman2020,otsubo2020,kim2020} and thermodynamic-uncertainty bounds~\cite{barato2015}.
    \item \textbf{Irreversibility and memory:} Whether memory can itself source irreversibility remains open. We have shown this requires a breaking of the fluctuation--dissipation relation, which the present construction does not contain and which is left to separate work. The long-range scaling of the nonlocal action, which would organize a field-theoretic and renormalization-group treatment, likewise lies beyond the present scope.
    \item \textbf{Scope and attribution (limitation):} While we attribute the non-Markovian character to attention, other components of a deep network, the residual stream, feedforward blocks, and normalization, also accumulate information across depth and could contribute to the effective memory kernel. Disentangling these from attention's contribution remains an empirical question.
\end{enumerate}
This paper establishes the foundation on which those rest: the local action is not an independent postulate, but the short-memory limit of a nonlocal path measure.

\section{Conclusion}
\label{sec:conclusion}
An attention-driven world model is fundamentally nonlocal. Its predictive path measure is the projection of a hidden linear-Markov augmentation, with memory entering as a nonlocal quadratic form in an Onsager--Machlup action whose kernels belong to the completely monotone class. The local Onsager--Machlup action of the companion paper is not an independent assumption but the short-memory limit of this nonlocal measure; notably, the irreversible channel identified in the companion paper survives this projection intact. This recovery is controlled for the reversible kinetic sector, which allows for a term-by-term short-memory expansion. The local action's Jacobian term is instead fixed by the convention through which the augmented representation transitions to a genuinely Markovian dynamics, and we have not treated it as a series expansion in $\eps$. In short, locality is a derived limit rather than a starting point.

This reframes how the predictive object should be interpreted. Treating the world model as a measure over paths, rather than a one-step conditional, redefines its memory as a structural feature of the action with a definite scaling in $\eps$. It also transforms the locality assumed in earlier analyses into a controlled approximation with explicit corrections. The representation result provides a precise characterization of this: within the linear passive class, asking whether a latent dynamics is nonlocal is equivalent to asking whether it masks a hidden Markov augmentation, and the completely monotone kernels mark exactly where the two coincide. Because the leading-order action is dimension-independent, the recovery of the local theory is an intrinsic property of the mechanism, not an artifact of the specific solvable example used for validation. The representation is also falsifiable: if the memory kernel estimated from a trained model fails to be completely monotone, for instance if its spectrum carries an oscillatory component, the dynamics lies outside the linear positive-spectrum projection assumed here, and the short-memory reduction to the local action need not hold. Locality, as a derived limit, is thus a statement that can be refuted by measurement.

Within the larger program, this paper supplies the bridge between a local effective description and the underlying nonlocal dynamics. It leaves the reversible and irreversible sectors cleanly separated, making the next step well-posed: once memory is understood to renormalize only the reversible metric, a memory-sourced contribution to irreversibility can be isolated as a genuine fluctuation--dissipation violation, rather than being conflated with the kernel itself. The measurement of these quantities in trained models, and the scaling analysis that would extend the construction toward a field theory, now rest on a foundation where the central objects, the path measure and its short-memory limit, are already firmly established.

\end{document}